\documentclass[12pt,preprint]{aastex}

\shortauthors{Zhou et al.}

\begin{document}

\title{A narrow line Seyfert\,1--blazar composite nucleus in 2MASX 
       J0324+3410}

\author{Hongyan Zhou\altaffilmark{1,2,3}, Tinggui Wang\altaffilmark{1,2}, 
Weimin Yuan\altaffilmark{4},
Hongguang Shan\altaffilmark{4},  Stefanie Komossa\altaffilmark{5},
Honglin Lu\altaffilmark{1,2}, Yi Liu\altaffilmark{6}, 
Dawei Xu\altaffilmark{7}, J. M. Bai\altaffilmark{4}, 
D.R. Jiang\altaffilmark{6}
}

\altaffiltext{1}{Center for Astrophysics, University of Science
and Technology of China, Hefei, Anhui, 230026, P.R.China}

\altaffiltext{2}{Joint Institute of Galaxies and Cosmology, SHAO
and USTC}

\altaffiltext{3}{Department of Astronomy, University of Florida,
Gainesville, FL 32611i, USA}

\altaffiltext{4}{National Astronomical Observatories/Yunnan
Observatory, Chinese Academy of Sciences, Kunming, Yunnan, P.O.
BOX 110, P.R.China}

\altaffiltext{5}{Max-Planck-Institut f\"{u}r extraterrestrische
Physik, Postfach 1312, 85741 Garching, Germany}

\altaffiltext{6}{Shanghai Astronomical Observatory, Chinese Academy of 
Sciences, Nandan Road, Shanghai, China}

\altaffiltext{7}{National Astronomical Observatories, Chinese
Academy of Sciences, A20 Datun Road, Chaoyang District, Beijing
100012, China}

\email{twang@ustc.edu.cn}

\begin{abstract}
We report the identification of 2MASX J032441.19+341045.9
(hereafter 2MASX J0324+3410) with an appealing object 
which shows the dual properties of both a narrow line 
Seyfert 1 galaxy (NLS1) and a blazar. 
Its optical spectrum, which has 
a $H\beta$ line width about 1600~km~s$^{-1}$ (FWHM), 
an [OIII] to $H\beta$ line ratio 
$\backsimeq 0.12$, 
and strong FeII emission, 
clearly fulfills the conventional definition of NLS1s.  
On the other hand, 2MASX J0324+3410 also exhibits some
behavior which is characteristic of blazars, including
a flat radio spectrum above 1 GHz, 
a compact core plus a one-sided jet structure on mas-scale 
at 8.4 GHz, 
highly variable fluxes in the radio, optical, and X-ray bands,
and a possible detection of TeV $\gamma$-ray emission.
On its optical image, obtained with the HST WFPC2,
the active nucleus is displaced from the center of the host galaxy,
which exhibits an apparent one-armed spiral structure extended to 
16 kpc.
The remarkable hybrid behavior of this object presents a challenge to 
current models of NLS1 galaxies and $\gamma$-ray blazars.

\end{abstract}


\section{Introduction}

Since their identification as a special subgroup of broad line
active galactic nuclei (AGNs), 
narrow line Seyfert 1 galaxies (NLS1s, Osterbrock \& Pogge
1985) have drawn substantial attention in the AGN
community over the last twenty years. 
The conventional definition of NLS1s consists of
two criteria: 
1) a narrow width of the broad Balmer emission line
($FWHM(H\beta)<2\,000$~km~s$^{-1}$); 
and 2) weak forbidden lines ($[OIII]\lambda5007/H\beta<3$). 
Subsequent studies revealed their
other unusual properties, such as 
1) strong FeII multiplet emission (e.g.,
Boroson \& Green 1992; Grupe et al. 1999; Veron et al. 2001; 
Zhou et al. 2006); 2) steep soft X-ray spectra (Wang et
al. 1996; Boller et al. 1996; Grupe et al. 1998); 3) rapid and large amplitudes of X-ray
variability (e.g., Leighly 1999; Komossa et al. 2000); 
and 4) commonly blue shifted UV line profiles (e.g.,
Leighly \& Moore 2004). 
These unusual properties are in fact an
extension to the extremity of a set of correlations between
H$\beta$ line width and the other observables. 
These correlations form the so-called eigenvector 1 (E1, Boroson \&
Green 1992; Sulentic et al. 2000). E1 is considered as the manifestation 
of the variation of one or a certain combination of several fundamental 
parameters of AGNs, such as the black hole mass and/or accretion rate.  

NLS1s are usually radio-quiet (RQ) with radio loudness $R<10$
(defined as radio to B-band optical flux ratio, $R\equiv f_{5GHz}/f_{B}$). 
The last decade has witnessed the identification of about two dozen
radio-loud (RL) NLS1s 
(Grupe et al. 2000; Zhou \& Wang 2002; Zhou et al. 2003; Zhou et al. 2005; 
Komossa et al. 2006a,b; Zhou et al. 2006). 
Given the intriguing fact that RL AGNs and ``normal'' NLS1s 
occupy the opposite extremes of
the E1 parameter space (Boroson 2002), the study of RL NLS1s 
may provide important clues to understanding 
the physical drivers of E1.

First proposed by Spiegel in 1978 to refer to 
rapidly variable AGNs, 
blazars, including BL Lac objects and flat spectrum radio quasars 
(FSRQ), are another small distinct subset of AGNs. 
An operational definition is: flat radio spectrum 
above 1 GHz, fast variability, high and variable polarization,
superluminal motion, and high brightness temperature (Urry \& Padova 1995). 
Blazars are believed to be radio loud AGN viewed 
(almost) along the direction of their radio jets; 
so that non-thermal jet emission is relativistically boosted.  
In our recent work, 
we found that the two radio loudest NLS1s, SDSS
J094857.3+002225 (Zhou et al.\ 2003) and 0846+51W1 (Zhou et al.\
2005), both  show blazar-like behavior. 
These results suggest that relativistic beaming 
may play an important role in very radio loud (VRL) NLS1s.

In this letter, we report the discovery of 
an appealing NLS1--blazar composite object, 
J032441.19+341045.9 (hereafter 2MASX J0324+3410),
an extreme and the best example so far
in the line of the NLS1-blazar connection.
This object was first observed spectroscopically 
by Remillard et al. (1993) as an
optical counterpart of an X-ray source detected in the HEAO-1
X-ray survey, and was classified as a Seyfert 1 galaxy.
It was subsequently observed by March$\tilde{\rm a}$
et al. (1996) but was (mis-)classified 
as a narrow line radio galaxy (NLRG). 
Here we present our new spectroscopic observation and
classification, along with its broad band property; 
we defer to a future paper for  
detailed analyses of its full data set from 
both archives and our on-going monitoring observations.
Throughout this paper, we assume a
cosmology with $H_{0}$= 70 km\, s$^{-1}$\,Mpc$^{-1}$,
$\Omega_{M}=0.3$, and $\Omega_{\Lambda}=0.7$.

\section{Observations, data analysis, and broad band properties}

\subsection{Optical Spectroscopy and NLS1 classification} 

We took spectra of 2MASX J0324+3410 with the
OMR spectrograph attached to the Cassegrain focus of
the 2.16m telescope at Xinglong Station of 
National Astronomical Observatory of China (NAOC) 
on Nov 25, 2005. A 1200 line mm$^{-1}$
grating (50 \AA~ mm$^{-1}$ dispersion) and a Tek
$1300\times1024$ CCD were used to cover a wavelength range of 1200 \AA~ 
centered at 5100\,\AA. 
Four exposures of 60 minutes each
were taken. KPNO standard stars were observed for
absolute flux calibration. 
A slit width of $2\farcs 5$ was chosen to match the seeing disk. 
The spectral resolution as measured from
the night-sky lines was 2.75\AA~ at FWHM. The CCD reduction,
including bias subtraction, flat-field correction, and cosmic-ray
removal were accomplished following the standard procedures using IRAF. 
The spectra were extracted and combined into a 1-D spectrum 
with the Galactic extinction $E(B-V)=0^{m}.213$ corrected.

\begin{figure}
\plotone{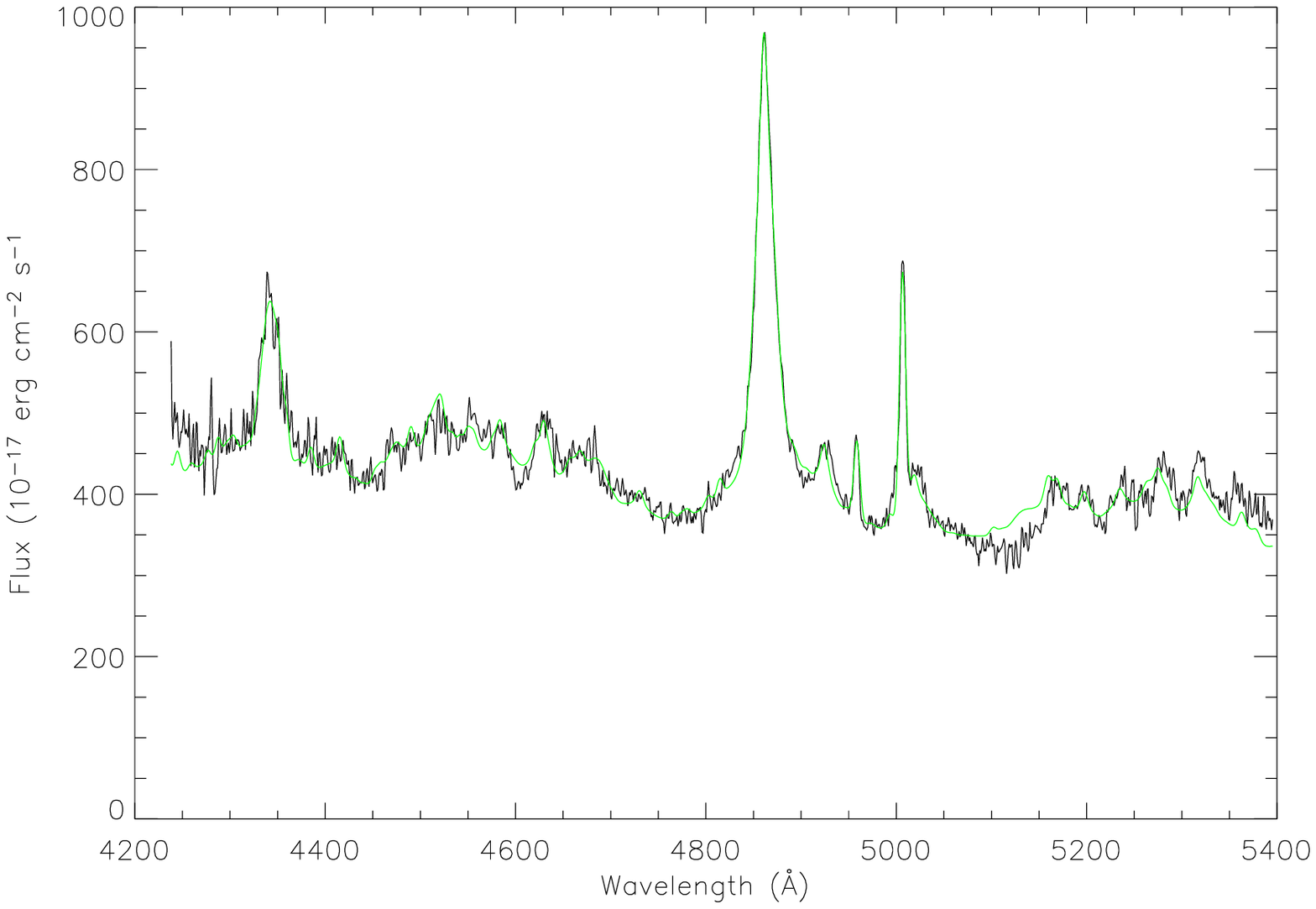} 
\caption{Observed spectrum of 2MASX
J0324+3410 and the best fit model (the green line).} \label{f1}
\end{figure}

We fit the spectrum with the following model components: 
1) a power law continuum; 
2) two Gaussians for the [OIII]$\lambda\lambda$4959, 5007 doublet; 
3) two Lorentzians for H$\beta$ and H$\gamma$, respectively; and 
4) the optical FeII multiplets of V$\acute{e}$ron-Cetty et al. (2004). 
The redshifts and profiles of the [OIII] doublet are taken to be the same,
and their flux ratio is fixed to the theoretical value. 
We also assume that the broad component of H$\beta$, H$\gamma$, and  
the FeII multiplets have the same redshifts and line 
profiles; and the same assumption is for the narrow component of  
these lines\footnote{V$\acute{e}$ron-Cetty's FeII model includes forbidden 
FeII lines.}. 
The result is displayed in Figure \ref{f1}. 
We also tried to fit 
the broad component of the Balmer lines with Gaussian,
and found that they
cannot be fitted with a single
Gaussian but two Gaussians are needed.
This latter model yields somewhat broader FWHM
but similar line strength 
compared to the former Lorentzian model
for both the FeII and Balmer lines.

The redshift of 2MASX J0324+3410 as determined from [OIII] is
$z=0.0629\pm0.0001$, consistent with that of the Balmer lines within
the measurement uncertainties. The Balmer line width,
FWHM = 1\,520~km~s$^{-1}$ for a Lorentzian model and 1\,650~km~s$^{-1}$ 
for a double Gaussian model (after correction for instrumental broadening), 
is narrower than that of normal Seyfert 1 galaxies and quasars,
but typical of NLS1s. 
The line ratio $[OIII]\lambda5007/H\beta$ is  $\simeq 0.12$
and is almost independent of the choice of the line profiles.
This value indicates that the bulk of the Balmer emission lines do not
originate from the narrow line region (NLR) and hence excludes the
possibility that 2MASX J0324+3410 is a type 2 AGN, e.g. a NLRG,
as was claimed in March$\tilde{\rm a}$ et al. (1996). 

The optical FeII complexes are rather strong with
$EW(FeII\lambda4570)\sim 130\AA$ and $R_{4570}\equiv
FeII\lambda4570/H\beta\sim 2.0$, where the FeII blend is
integrated from 4434 to 4684 \AA~  for both broad and narrow
components with larger EW(FeII) and $R_{4570}$ for the Lorentzian model.
Remillard et al. (1993) reported a $R_{4570}=1.54$.
In summary, our optical spectrum clearly
shows that 2MASX J0324+3410 is a classic NLS1.

\subsection{Broad Band Properties and SED}

Radiation from 2MASX J0324+3410 was detected in a wide
wavelength range across almost the whole electromagnetic spectrum,
in the radio, infrared (IRAS and 2MASS), optical band (e.g., Swift UVOT), 
X-ray and possibly $\gamma$-ray bands. 
It is a strong radio source with a flux density 304$-$581 mJy at 5 GHz,
and shows a flat radio spectrum up to at least 10 GHz and 
significant flux variations. 
Using simultaneous observations at 2.695, 4.75, and 10.55 GHz,
Neumann et al. (1994) found a 
flat spectral index, $\alpha\simeq 0.1$, and
detected polarization at the 3\%, 5\%, and 4\% level,  respectively.
A factor of 1.3 variation ( 4$\sigma$ level) 
in its 1.4 GHz flux on time-scales of ten years was found
by comparing the
Green Bank 1.4 GHz Northern Sky Survey (GBNSS, White \& Becker 1992) and 
the NRAO VLA Sky Survey (Condon et al. 1998). 
Even larger amplitude  variations (a factor of 1.6 and 1.9) 
on shorter time scales (3.5 and 3.3 yrs) were 
observed at 5 GHz (Neumann et al. 1994; Griffith et al. 1991; 
Laurent-Muehleisen et al. 1997). 
These variations are significant at the 
3--6 $\sigma$ level, and cannot be attributed to source contamination
as the beam sizes for observations at the higher flux states are
(fortunately) smaller than those at the lower states. 
Assuming the flux variability is intrinsic, we can derive a lower limit 
on the brightness temperature of $5\times 10^{11}~{\mathrm K}$ (c.f., Zhou 
et al. 2006). 
This value is an order of magnitude higher than the
equipartition value $\sim 5\times 10^{10}$ K (e.g., Readhead
1994)
but close to the inverse Compton
limit of $10^{12}$ K (Kellermann \& Pauliny-Toth 1969). 

The source was also observed by the VLBA in the VLBA calibrator survey (VCS). 
High resolution images at 2.2 GHz and 8.4 GHz reveal a core and a 
weak extended south-west component 6.8 mas (8.2 pc) away from the 
core (Beasley et al. 2002). 
We have re-processed the VLBA data. 
The core is further resolved into a brightest 
component in the north-east and an extension towards the south-west,
reminiscent of a core--jet structure. 
The brightest compact component has a deconvolved 
size of about 0.1 mas (0.12 pc) and a flux of 0.165 Jy at 8.4 GHz,
leading to a brightness temperature of about 4$\times 10^{11}$ K, 
consistent with the estimate from the variability analysis above. 
The south-west component has a steep radio spectrum and
seems to  have no proper motion ($<0.05c$; Yuan W. et al. in 
preparation), and therefore is possibly a weak radio lobe.

The radio loudness of 2MASX J0324+3410 is estimated to be 
$R=f_{5GHz}/f_{\mathrm B}= 38-71$ using the HST magnitude and 
assuming $\alpha_{opt}=0.5$ for the nucleus. 
If we adopt the swift UVOT B magnitude of B=16.11 mag, we obtain
$R\simeq$89-151, after correction for Galactic extinction. Note 
that the Swift B magnitude may include some contamination of the 
galactic light (see last section), so there can be a significant 
variation between the Swift and HST observations. Since the 5 GHz 
radio emission is likely
subject to substantial enhancement due to the Doppler beaming effect,
we try to estimate its `intrinsic radio loudness' using
a low-frequency flux, which is thought to be less 
affected by Doppler boosting.
Assuming typical spectral indices $\alpha_r=0.7-1.0$ for the extended radio 
components, we find an `intrinsic radio loudness' 4--25 from the radio 
flux 1.02 Jy at 151 MHz (Hales, Baldwin \& Warner 1993). 
This puts 2MASX J0324+3410 in the class of radio intermediate quasars. 
Its radio power, at 178 MHz, $P_{178{\mathrm MHz}}\lesssim 
8\times 10^{24}~{\mathrm W~Hz}^{-1}$, as interpolated from the one at 
151 MHz, is below the dividing line separating the FR~II and FR~I types. 

X-ray emission as well as hard X-ray (and perhaps even $\gamma$-ray) 
flares have been detected from 2MASX J0324+3410 by various instruments.
For observations with ROSAT and Swift,
we reduced the publicly available
data to derive the X-ray spectra and fluxes; 
for the rest of the data sets (mainly sky monitor type instruments), 
we made use of the X-ray count rates provided as available from NED. 
Hard X-ray emission was detected 
with the HEDs (High Energy Detectors, 2.6--60~keV) on-board HEAO-1
at three epochs over a time span of one year,
when it showed decreasing averaged count rates
from  $1.25\pm 0.39~cts~s^{-1}$ to $0.01\pm 0.26~cts~s^{-1}$.
It was also detected in the ROSAT All Sky Survey (RASS), 
with a 0.1--2.4\,keV flux of 
3.1$\pm0.5\times 10^{-12}$~ergs~cm$^{-2}$~s$^{-1}$ in 
the observed frame and corrected for the Galactic absorption 
($N_H^G=1.45~10^{21}$~cm$^{-2}$, (Dickey \& Lockman 1990). 
2MASX J0324+3410 has been monitored 
by the All Sky Monitor (ASM) 
on-board Rossi X-ray Timing Explorer (RXTE) since June 1996. 
The average ASM count rate over the last ten years is 0.0343$\pm$0.0065 cts~s
$^{-1}$, converting to a 2--10\,keV flux of $F_X=9.0\times10^{-12}$ 
ergs~cm$^{-2}$~s$^{-1}$.  
The most recent X-ray observation was done by the Swift X-ray
Telescope (XRT) in July  2006.  
The spectrum in the 0.3-10~keV band can be well  
fitted with a power-law of a photon index 2.02$\pm$0.06 
with Galactic absorption. 
X-ray flux variations up to a factor of two on 
time-scales less than 1~ks have been detected in the Swift data.
The average flux in the 0.2--2.4 keV band was
3 times brighter than that measured in the RASS. 
Detailed analysis of the X-ray data-sets will be presented 
in a future paper (Yuan W. et al.
in preparation). Of particular interest,
a TeV flare was claimed to be marginally detected 
at a significance level of 
$\sim 2.5-3.3~\sigma$ on October 10, 2001 with the Whipple imaging air
Cerenkov telescope at the Whipple Observatory (Falcone et al. 2004).
The average $\gamma$-ray count rate was $0.46\pm 0.14$ times that
of the Crab, with a peak rate $0.62\pm 0.19$ Crab. 

\begin{figure}
\plotone{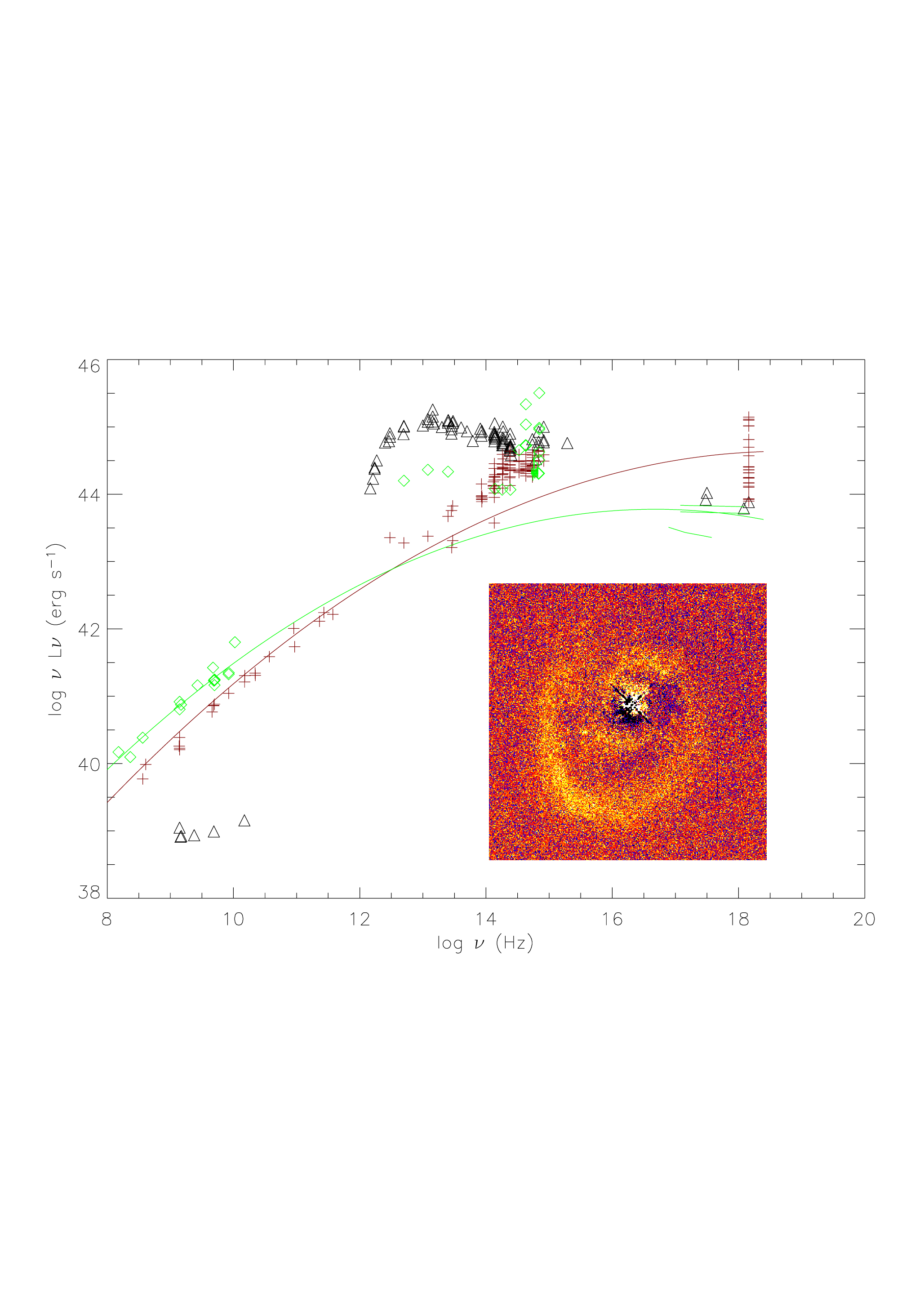} 
\caption{Broad band spectral energy
distribution of 2MASX J0324+3410 (green diamonds). A typical NLS1, 
IZw 1 (black triangles), and a typical TeV blazar, Mrk 421 (cross), are
also plotted for comparison. The 2-order polynomial (parabolic)
fit of the radio and X-ray data of 2MASX J0324+3410; and of the radio, IR,
and X-ray data of Mrk 421 is shown as dashed and dotted line, 
respectively. Data apart from those described in the text were retrieved 
from the HEASARC browser server (http:\/\/heasarc.gsfc.nasa.gov\
/cgi-bin\/W3Browse.). 
Inset: a residual image after subtraction of the best-fit 
model of a point source plus a Sersic component from the HST 
WFPC image. The similarity with a one-armed spiral-like structure 
can be clearly seen.}   
\label{f2}
\end{figure}

The broad band spectral energy distribution
(SED) for the nucleus of 2MASX J0324+3410
is plotted in Figure \ref{f2},
using the results of our own data as well as 
data collected through the HEASARC web browser.
For comparison, the SED for IZw 1 and Mrk 421---a 
well known NLS1 and blazar, respectively--are
over-plotted.
It can be seen that 2MASX J0324+3410 resembles 
Mrk 421 in terms of the broad band, non-thermal continuum
in the radio and possibly X-ray bands,
whereas it resembles IZw 1 in the infrared--optical regime
where thermal emission is dominant.

\subsection{The host galaxy --- a one-armed spiral?}

Two snapshot images of 2MASX J0324+3410, each of 200\,s exposure,
were taken with the second
Wide Field and Planetary Camera (HST/WFPC2) of 
the Hubble Space Telescope (HST) with the 
F702W filter ($\lambda_{eff}=6919~\AA$).
The data were retrieved from the HST archive.
The two exposures were combined to create a single image
with a better S/N ratio.
A ring-like structure of $\sim 15^{''}$  diameter
can be clearly seen, which corresponds to $\sim 15.6$~kpc
at redshift 0.0629.
A nuclear source with high  surface brightness
is apparently displaced from the symmetric center of the host galaxy.
Thanks to the superb spatial resolution of  $\lesssim 0^{''}.1$ of HST,
we were able to extract  the surface brightness profile of the 
galaxy and decompose it into two components:
an unresolved point-like source 
as the NLS1 nucleus, and a S{\'e}rsic model for the host galaxy   
(detailed analysis is to be presented in a later paper).
We find $m_{NLS1,F702}=15.15 $~mag for the active nucleus, and 
$m_{bulge,F702}=15.22$~mag for the host galaxy. 
After correcting for the Galactic extinction E(B-V)=0.213 Mag
we get $m_{NLS1, F702}\thickapprox 14.70$~mag. The residual looks 
like a one-armed spiral (see Fig 1b).  

\section{Discussion: a NLS1--blazar composite}

The spectral and temporal properties of  2MASX J0324+3410 
in the radio and X-ray
bands, as presented in Sect.~2.2, 
are characteristic of blazars:
a flat radio spectrum and flux variability, 
a compact and bright radio core on milli-arcsec
scale, hard X-ray emission and flares, 
short timescale variations in X-rays,
and a broad band, non-thermal continuum.
In particular, its SED and possible TeV $\gamma$-ray emission
resemble those of a high-energy peaked blazar (HBL). 
On the other hand, its optical properties fulfill all criteria for
   classification as NLS1.
These observational facts thus make 2MASX J0324+3410
a composite of a NLS1 and a blazar.
We note that the rest frame H$\beta$ 
equivalent width $EW(H\beta)=58\pm 4$\AA~ is only 
slightly below the median value ($\backsimeq 80~\AA$) for a large sample 
of NLS1s (Zhou et al. 2006), i.e. the contribution from a potential 
non-thermal continuum is small in the optical.
If the X-ray to optical spectral slope is similar to that of
I ZW 1 for the NLS1 component, the NLS1 accounts for only
$\sim$1/3 of the observed X-ray emission, which also has a 
flat spectrum compared with other NLS1s.
Its radio and most likely X-ray radiation can be naturally
explained as being from a jet (via synchrotron emission),
while the  infrared and optical light is dominated by
thermal emission from a Seyfert nucleus.

We estimate the central black hole (BH) mass using a few methods.
Using the width and luminosity of the H$\beta$ line and the
empirical scaling relations as in Greene \& Ho (2005), we find
a BH mass of 10$^7 M_{\mathrm \odot}$. While the empirical
relations of Vestergaard \& Peterson (2006) give BH masses of 
3$\times10^7 M_{\mathrm \odot}$ 
using the continuum luminosity at 5100\AA and 
1.8$\times 10^7 M_{\mathrm \odot}$ using the H$\beta$ luminosity.
These BH mass estimates\footnote{
We note that a higher black hole mass of 
$8\times10^7 M_{\mathrm \odot}$ 
is estimated if one uses the line dispersion instead of the 
{\it FWHM}, following Collin et al. (2006).
} are consistent within their 
uncertainties, giving $M_{BH}\sim 10^{7}~M_{\odot}$.
Interestingly, this value falls into the overlapping region
in the BH mass distributions for NLS1s and blazars,
which have the bulk lying within $10^{6-7}~M_{\odot}$ 
(e.g., Wang \& Lu 2001; Grupe \& Mathur 2004; Zhou et al. 2006)
and $10^{7-9}~M_{\odot}$ 
(e.g., Woo et al. 2005, Falomo et al. 2002), respectively.

We estimate the bolometric luminosity from the 5100\AA 
luminosity using the correction factor 
for quasars given by Elvis et al. (1994).
This gives $L_{\rm bol}\thickapprox 1.2~10^{45}~erg~s^{-1}$.
We thus obtain a rough estimate of the Eddington ratio  
$\dot{m}\equiv L_{\rm bol}/L_{\rm Edd}\thickapprox 0.1$
for 2MASX J0324+3410.
The Eddington ratio is typical of NLS1s,
and resembles FSRQ as far as its blazar property is concerned;
however, the SED of its jet emission is more like  
HBLs, which have, on the contrary,
generally low accretion rates.
The one-armed spiral may provide such fueling to the 
active nucleus, though
it cannot be completely ruled out that the HST image we see is
a structure made up by dust lanes or even a ring galaxy.

2MASX J0324+3410 is not unique, 
but is the most representative object of its kind,
which seem to be rare.
The optical spectrum of the BL Lac object 0846+51W1 
showed typical NLS1 characteristics at low states (Zhou et al. 2005). 
The radio-to-optical SED of the radio loud NLS1 
J094857.3+002225, for which beaming is required to explain the high brightness 
temperature, is also similar to that of 2MASX J0324+3410  (Zhou et al. 
2004; also Doi et al. 2006).  
A third example is RXJ 16290+4007 (Zhou \& Wang 
2002; also Komossa et al. 2006b), 
a flat-spectrum radio quasar with NLS1 characteristics in the optical.
Furthermore, 8 out of the 9 very-radio-loud 
(VRL, $R_{1.4}\equiv f_{1.4GHz}/f_{B}\gtrsim 250$) NLS1s
currently under study  (Zhou et al. 2006b in preparation) 
show flat radio spectra: 
\footnote{
the only exception so far is 
SDSS~J172206.03+565451.6 with $\alpha_{0.33-1.4GHz}=0.7$ 
(Komossa et al. 2006b).
},
 $\alpha_{1.4-5GHz}
\lesssim 0.5$, $f_{\nu}\propto\nu^{-\alpha}$, which is remarkable. 
Thus, it is probably true that most VRL NLS1s are actually 
radio intermediate (RI) NLS1s with boosted jet emission (Zhou et al. 
2006b). If so, they are similar to a population of boosted RI quasars 
recently identified by Wang et al (2006) from their high radio 
brightness temperature, but with more extreme optical properties.  
They may be an analog to the radio-bright, very high soft states in 
black hole binaries (e.g., Fender, Belloni \& Gallo 2004).

The hybrid state of 2MASX J0324+3410 raises questions to the current 
understanding of TeV/high-energy peaked blazars. 
It is widely accepted that this type of blazar have 
very low accretion rates, but 2MASX J0324+3410 is 
 doubtlessly an exception. A common notion is that the intensive radiation 
field of the accretion disk prevents electrons from gaining the high energies 
that are required for TeV $\gamma$-ray production or the formation of high 
frequency synchrotron peaks in the SED. This is certainly not the case for 
2MASX J0324+3410.    

\acknowledgments We thank Jianyan Wei and Jing Wang for help in
our spectroscopic observations, and Dirk Grupe and Binbin Zhang in the 
analysis of the swift data. This work was supported by Chinese
NSF through NSF10233030, NSF10533050 and NSF10473013, and the 
Bairen Project of CAS. D.W. Xu acknowledges the Chinese NSF support 
through NSFC10503005. This paper has made use of data from NED.

\clearpage

\end{document}